\documentclass[floatfix,twocolumn,showpacs,aps,tightenlines]{revtex4-1}
\usepackage{graphicx}
\usepackage{color}
\usepackage{psfrag}
\usepackage{dcolumn}% Align table columns on decimal point
\usepackage{bm}% bold math

\usepackage{amssymb}
\usepackage{amsmath}
\usepackage{amsfonts}
\usepackage{longtable}
\usepackage{xspace}

\newcommand{\beq}{\begin{equation}}
\newcommand{\eeq}{\end{equation}}

\newcommand{\be}{\begin{equation}}
\newcommand{\ee}{\end{equation}}
\newcommand{\bea}{\begin{eqnarray}}
\newcommand{\eea}{\end{eqnarray}}
\newcommand{\bes}{\begin{subequations}}
\newcommand{\ees}{\end{subequations}}

\usepackage{bm}

\begin{document}

\title{The RIT binary black hole simulations catalog}

\author{James Healy}
\author{Carlos O. Lousto}
\author{Yosef Zlochower}
\author{Manuela Campanelli}
\affiliation{Center for Computational Relativity and Gravitation,
School of Mathematical Sciences,
Rochester Institute of Technology, 85 Lomb Memorial Drive, Rochester,
New York 14623}

\date{\today}

\begin{abstract}
  The RIT numerical relativity group is releasing a public catalog of
  black-hole-binary waveforms. The initial release of the catalog
  consists of 126 recent simulations that include
precessing and non precessing systems with
mass ratios $q=m_1/m_2$ in the range $1/6\leq q\leq1$.
The catalog contains information about the initial data of the simulation, the
  waveforms extrapolated to infinity,
  as well as information about the peak luminosity and final remnant 
  black hole properties. These waveforms can be used to independently interpret
  gravitational wave signals from laser interferometric detectors and the remnant properties to model the merger of binary black holes from initial configurations.
\end{abstract}

 \maketitle

\section{Introduction}\label{sec:Intro}

The
breakthroughs~\cite{Pretorius:2005gq,Campanelli:2005dd,Baker:2005vv}
in numerical relativity allowed  numerical relativists  to make
detailed predictions for the gravitational waves from the latest
inspiral, plunge, merger and ringdown of black hole binary systems
(BHB).  Numerical relativity predictions were  confirmed by the first
direct detection \cite{TheLIGOScientific:2016wfe} of gravitational
waves from such binary systems
\cite{Abbott:2016blz,Abbott:2016nmj,TheLIGOScientific:2016pea}
and by its comparison to targeted runs
\cite{Abbott:2016apu,Lovelace:2016uwp}.  Those observations are
consistent with general relativity as the correct theory for gravity
as discussed in
\cite{TheLIGOScientific:2016src,TheLIGOScientific:2016pea}.

The RIT group has been using
numerical relativity techniques to explore the late
dynamics of spinning black-hole binaries, beyond the post-Newtonian
regime for many years. 
This includes
simulations of the first generic, long-term
precessing binary black hole evolution without any symmetry were 
performed in Ref.~\cite{Campanelli:2008nk}, where a detailed comparison with 
post-Newtonian $\ell=2,3$ waveforms was made, as well as
studies of the {\it hangup}, i.e. the role individual black hole
spins play to delay or accelerate their merger \cite{Campanelli:2006uy},
the determination of the magnitude and direction 
of the {\it recoil} velocity of the final merged black hole
\cite{Campanelli:2007ew,Campanelli:2007cga,Lousto:2011kp},
and the {\it flip-flop} of individual spins during the orbital phase
\cite{Lousto:2014ida,Lousto:2015uwa,Lousto:2016nlp}.
Other numerical simulations have also explored the corners of parameter space,
such as mass ratios $q=1/100$ in Ref.~\cite{Lousto:2010ut}, and
larger initial separations $R=100M$ in \cite{Lousto:2013oza}. 
And also near extremal $\chi=0.994$ spinning black hole binaries in 
\cite{Lovelace:2014twa} by the SXS collaboration.

There have been several significant efforts to coordinate numerical
relativity simulations to support gravitational wave observations.
These include the numerical injection analysis (NINJA) project
\cite{Aylott:2009ya,Aylott:2009tn,Ajith:2012az,Aasi:2014tra}, the
numerical relativity and analytical relativity (NRAR) collaboration
\cite{Hinder:2013oqa}, and the waveform catalogs released by the
simulating extreme spacetimes (SXS) collaboration~\cite{Mroue:2013xna}
and Georgia Tech.~\cite{Jani:2016wkt}.

The paper is organized as follows. Section \ref{sec:FN}
describe the methods and criteria for producing the numerical
simulations.
We next describe in Sec.~\ref{sec:Catalog} the use and content
of the data in the public catalog.
We conclude with a discussion in 
Sec.~\ref{sec:Discussion} of the use and potential extensions 
to this work to precessing binaries.

\section{Full Numerical Evolutions}\label{sec:FN}

The runs in the RIT Catalog were evolved using the {\sc
LazEv}~\cite{Zlochower:2005bj} implementation of the moving puncture
approach~\cite{Campanelli:2005dd} with the conformal
function $W=\sqrt{\chi}=\exp(-2\phi)$ suggested by
Ref.~\cite{Marronetti:2007wz}. In all cases we use the BSSNOK
(Baumgarte-Shapiro-Shibata-Nakamura-Oohara-Kojima) family
of evolutions systems~\cite{Nakamura87, Shibata95, Baumgarte99}.
For the runs in the catalog, we used a variety of finite-difference
orders, Kreiss-Oliger dissipation orders, and Courant factors~\cite{Lousto:2007rj,
  Zlochower:2012fk, Healy:2016lce}. Note that we do not upwind the advection terms.
All of these are
given in the metadata included in the catalog and the references associated
with each run. 

The {\sc LazEv} code uses the {\sc EinsteinToolkit}~\cite{Loffler:2011ay,
einsteintoolkit} / {\sc Cactus}~\cite{cactus_web} /
{\sc Carpet}~\cite{Schnetter-etal-03b}
infrastructure.  The {\sc
Carpet} mesh refinement driver provides a
``moving boxes'' style of mesh refinement. In this approach, refined
grids of fixed size are arranged about the coordinate centers of both
holes.  The {\sc Carpet} code then moves these fine grids about the
computational domain by following the trajectories of the two BHs.

We use {\sc AHFinderDirect}~\cite{Thornburg2003:AH-finding} to locate
apparent horizons.  We measure the magnitude of the horizon spin using
the {\it isolated horizon} (IH) algorithm detailed in
Ref.~\cite{Dreyer02a} and as implemented in Ref.~\cite{Campanelli:2006fy}.
Note that once we have the horizon spin, we can calculate the horizon
mass via the Christodoulou formula 
%\begin{equation}
${m_H} = \sqrt{m_{\rm irr}^2 + S_H^2/(4 m_{\rm irr}^2)}\,,$
%\end{equation}
where $m_{\rm irr} = \sqrt{A/(16 \pi)}$, $A$ is the surface area of
the horizon, and $S_H$ is the spin angular momentum of the BH (in
units of $M^2$).  In the tables below, we use the variation in the
measured horizon irreducible mass and spin during the simulation as a
measure of the error in computing these quantities, since the levels
of gravitational wave energy and momentum absorbed by the holes is orders
of magnitude smaller.  
We measure radiated energy,
linear momentum, and angular momentum, in terms of the radiative Weyl
Scalar $\psi_4$, using the formulas provided in
Refs.~\cite{Campanelli:1998jv, Lousto:2007mh}. However, rather than
using the full $\psi_4$, we decompose it into $\ell$ and $m$ modes and
solve for the radiated linear momentum, dropping terms with $\ell >
6$.  The formulas in Refs.~\cite{Campanelli:1998jv, Lousto:2007mh} are
valid at $r=\infty$.  We extract the radiated energy-momentum at
finite radius and extrapolate to $r=\infty$. We find that the new
perturbative extrapolation described in Ref.~\cite{Nakano:2015pta} provides the
most accurate waveforms. While the difference of fitting both linear and
quadratic extrapolations provides an independent measure of the error.
Studies of the finite difference errors, and
verification that the waveforms provided in this catalog are computed
at a numerical resolution in the convergence regime can be found in
the appendices of Refs.~\cite{Healy:2014yta} and ~\cite{Healy:2016lce}.
Although higher multipoles modes [beyond  $(\ell,m)=(2,\pm2)$] are
not provided in the catalog for the sake of simplicity, they are used
(typically up to $\ell=4$ or $\ell=6$) 
in the computation of radiative quantities such as the energy and
linear and angular momenta.

To compute the initial low eccentricity orbital parameters
we use the post-Newtonian techniques described in~\cite{Healy:2017zqj}.
To compute the numerical initial data, we use the puncture
approach~\cite{Brandt97b} along with the {\sc
TwoPunctures}~\cite{Ansorg:2004ds} thorn. 

We measure the distance between the two BHs using the {\it simple
proper distance} or SPD. The SPD is the proper distance, on a given
spatial slice,
between the
two BH apparent horizons as measured  along the coordinate line
joining the two centers. As
such, it is gauge dependent, but still gives reasonable results
(see Ref.~\cite{Lousto:2013oza} for more details).

\section{Catalog}\label{sec:Catalog}

\begin{figure}
  \includegraphics[angle=270,width=\columnwidth]{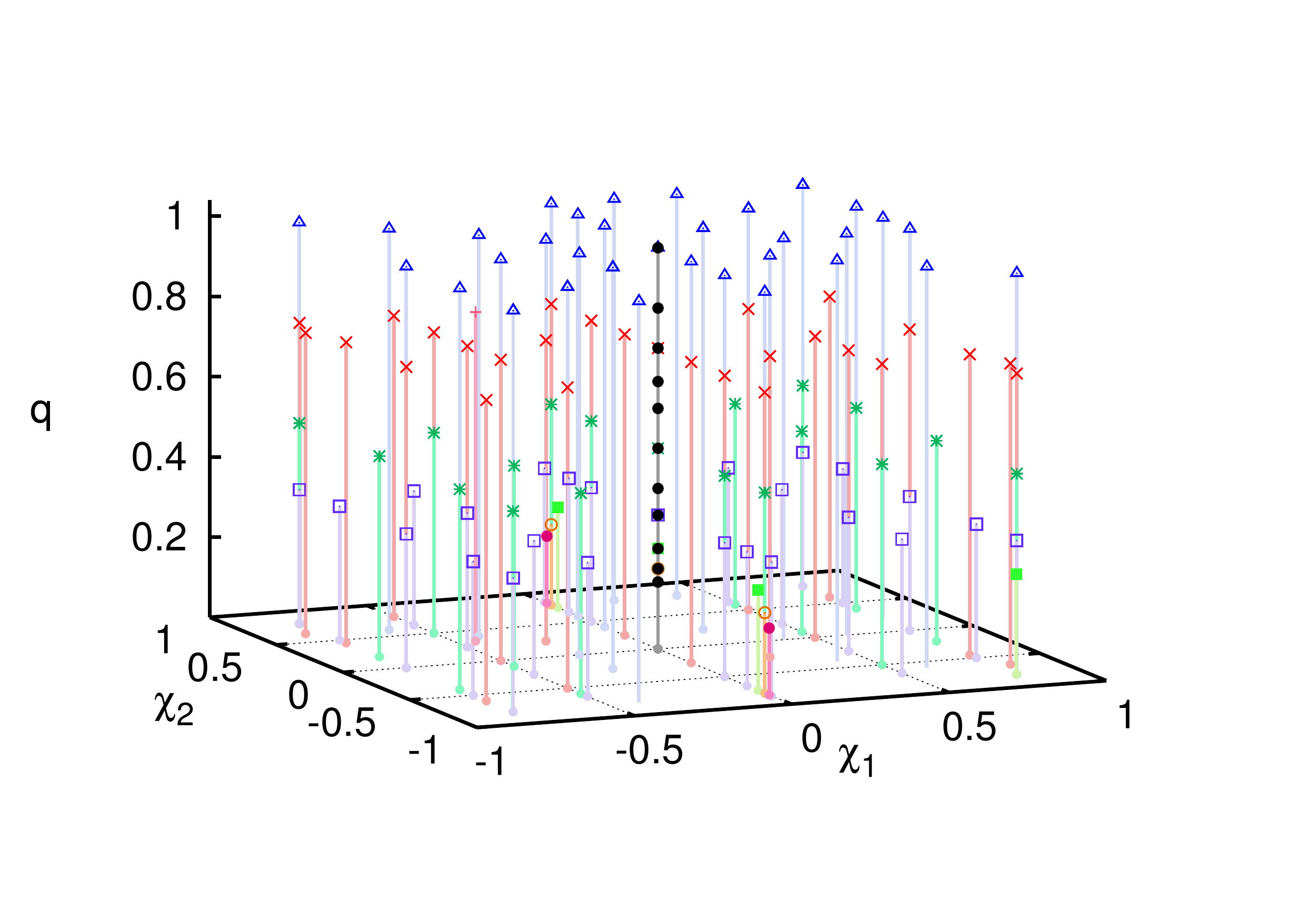}
  \caption{Initial parameters in the $(q,\chi_1,\chi_2)$ space
for the 120 nonprecessing binaries. Each mass ratio is given a different
color: blue triangles($q=1.00$), pink plus ($q=0.82$), red crosses ($q=3/4$), 
green stars ($q=1/2$),
purple squares ($q=1/3$), light green full squares ($q=1/4$), orange circles($q=1/5$), and 
brown full circles ($q=1/6$).  Nonspinning runs are black full circles.
  \label{fig:ID3d}}
\end{figure}

\begin{figure}
  \includegraphics[angle=270,width=\columnwidth]{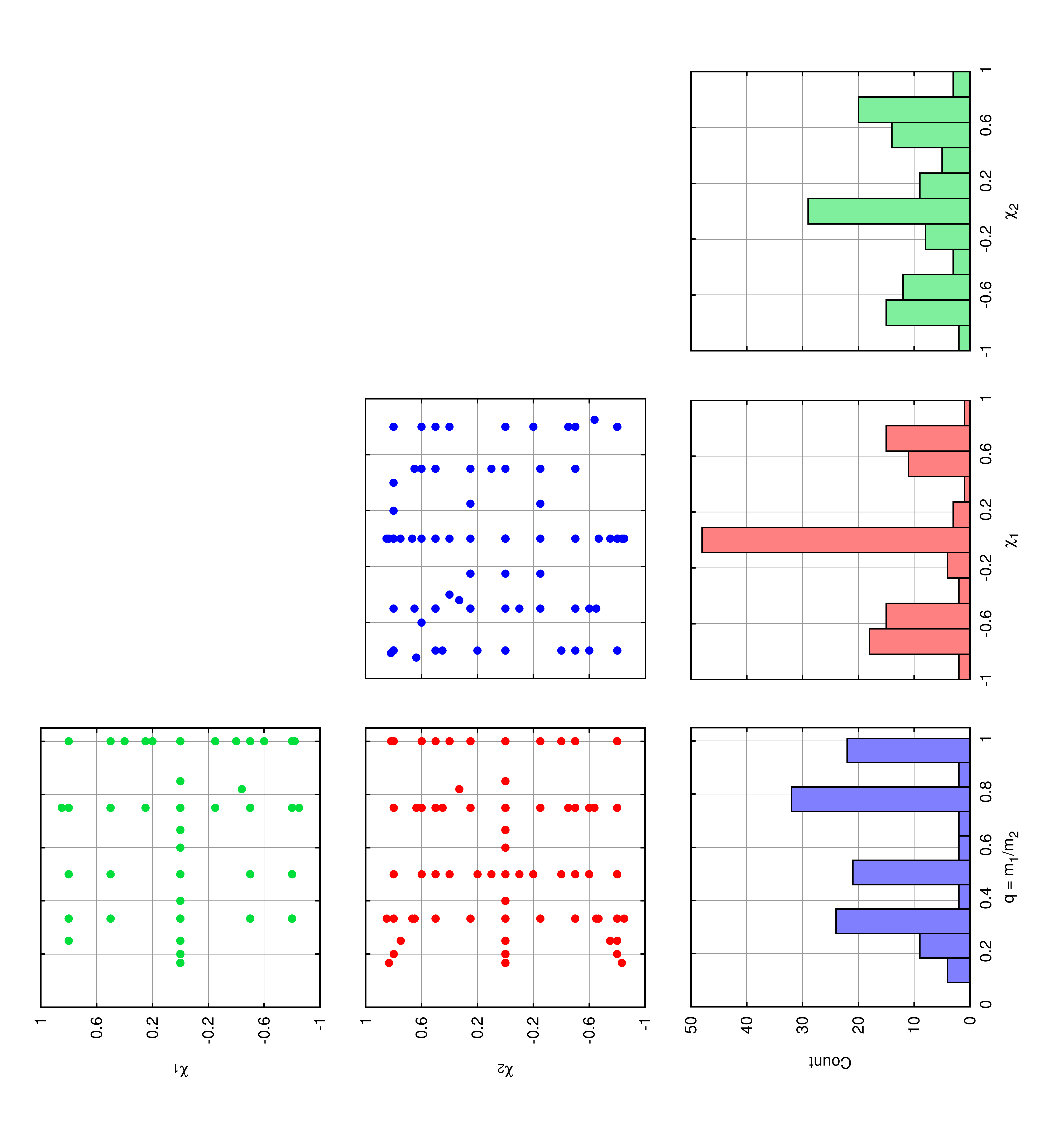}
  \caption{Initial parameters in the $(q,\chi_1,\chi_2)$ plane
for the 120 nonprecessing binaries. Note that here $q=m_1/m_2<1$.
  \label{fig:ID2d}}
\end{figure}

\begin{figure*}
  \includegraphics[angle=270,width=1.95\columnwidth]{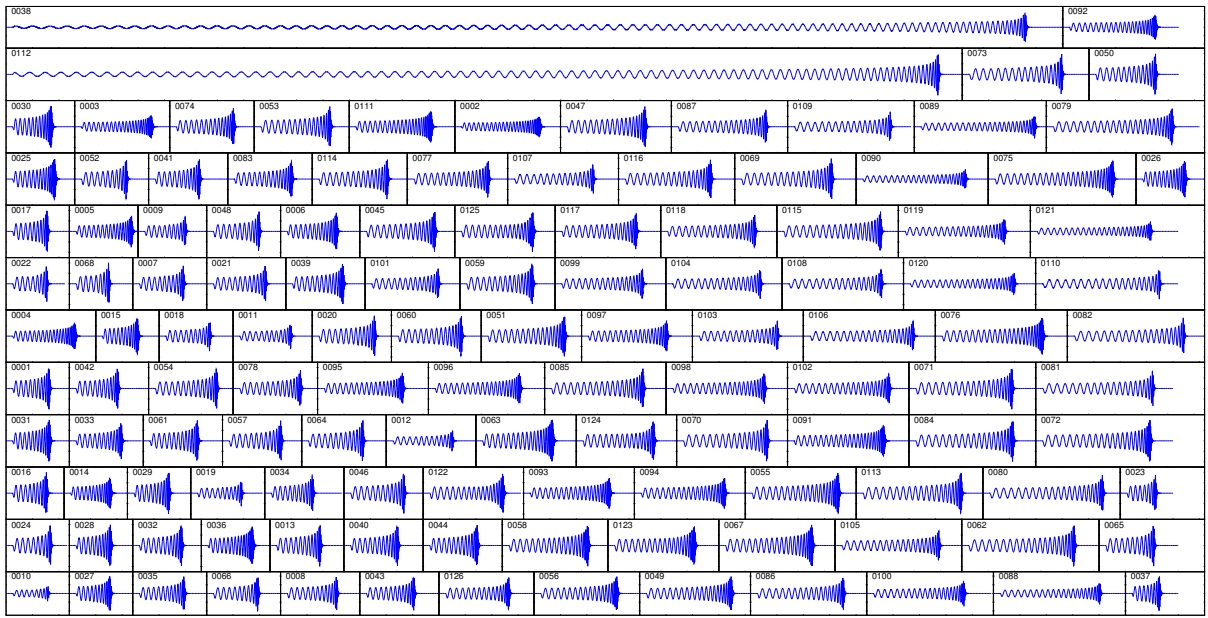}
  \caption{Visual display of the different lengths of the (2,2) waveforms
in this first delivery of 126 simulations in the RIT Catalog.  Each row
of waveforms spans ${\sim}22700 M$ of simulation time from edge to edge, 
with each tic mark denoting $500 M$.
  \label{fig:waveforms}}
\end{figure*}

The RIT Catalog can be found at \url{http://ccrg.rit.edu/~RITCatalog}.
Figure~\ref{fig:ID3d} shows the distribution of non-precessing runs in the catalog in
terms of $\chi_{12}$ and $q$ ($\chi$ here is the $z$-component of
the dimensionless spin).
The information currently in the catalog consists of the metadata
describing the runs and $(\ell=2, m=\pm2)$ modes of $r \psi_4$. 
The extrapolations of $r\psi_4$ to $r=\infty$  are performed using the perturbative approach
of~\cite{Nakano:2015pta}. The associated metadata include the
initial orbital frequencies, ADM masses, initial waveform frequencies,  black hole masses, momenta, spins, separations,
and eccentricities, as well the black-hole masses and spins once the
initial burst of radiation has left the region around the binary.
These {\it relaxed} quantities are more accurate for modeling purposes
than the initial masses and spins. In addition, we also include the
final remnant masses and spins.

The catalog is organized using an interactive table that includes an
identification number, resolution, type of run (nonspinning, aligned spins,
precessing), the initial simple proper distance between the two black
holes, the coordinate separation, the mass ratio of the two black
holes, the components of the dimensionless spins of the two black
holes, the starting waveform frequency, time to merger, number of
gravitational wave cycles, remnant mass, remnant spin, recoil
velocity, and peak luminosity. The final column gives the appropriate
bibtex keys for the relevant publications where the waveforms were
first presented. The table can be sorted (ascending or descending) by
any of these columns.

The initial waveform frequency, denoted by $M f_{22, {\rm start}}$ in
the table gives the starting frequency in units of 
$2.03\times10^5 \left(\frac{M_\odot}{M}\right)  {\rm Hz},$ where $M$ is the mass
of the binary and $M_\odot$ is one solar mass (e.g., $M f_{22, {\rm
start}} = 0.01$ corresponds to 34 Hz for a 60 $M_\odot$ binary). Note
that $2 \pi f_{22} = \omega_{22}$. The runs in the catalog span
initial frequencies from 0.003 to 0.012, with a corresponding initial
proper separations of 10.59M to 25.18M. Times from the start of the
simulation to merger range from 556M to 19 219M, and the number of
inspiral
cycles in the $(\ell=2, m=2)$ mode of $\psi_4$ range from 8.3 to 89.9.

Resolutions are given in terms of the gridspacing of the refinement
level where the waveform is extracted (which is typically two
refinement levels below the coarsest grid) with $R_{obs}\sim100M$.
We use the notation 
nXYY, where the gridspacing in the wavezone is given by M/X.YY, e.g.,
n120 corresponds to $h=M/1.2$.

For each simulation in the catalog there are two files: one contains
the metadata information in ASCII format, the other is a tar.gz file
containing ASCII files with the $(\ell=2, m=\pm2)$ modes of $r\psi_4$ (extrapolated to
$r=\infty$).

Note that our catalog provides the Weyl scalar $r \psi_4$ extrapolated
to infinity rather than the strain $\hat{h}$. We leave to the user to
convert $r\psi_4$ to strain (e.g., along the lines of the method
delineated in the NRAR collaboration ~\cite{Hinder:2013oqa}, for instance). 

Figure~\ref{fig:ID2d} shows the distribution of the 120 non-precessing runs in the catalog in
terms of $\chi_{1,2}$ and $q$ ($\chi$ here is the 
$L$ or $z$-component of
the dimensionless spin).

%\begin{widetext}
%\end{widetext}

\section{Conclusions and Discussion}\label{sec:Discussion}

The breakthroughs~\cite{Pretorius:2005gq, Campanelli:2005dd, Baker:2005vv}
in numerical relativity were instrumental in identifying the first detection of
gravitational waves \cite{TheLIGOScientific:2016wfe} with the merger
of two black holes. The direct comparison of numerical waveforms
with observations also allows one to determine the parameters of such
binary~\cite{Abbott:2016apu}. The current catalog of waveforms as
displayed in Fig.~\ref{fig:waveforms} can be used to
perform independent analysis by the wider gravitational wave community
and serves as a platform to deliver new sets of simulations as they
become available.

Aside the interest in producing waveforms for direct comparison
with observation, the simulations of orbiting black hole binaries produce
information about the final remnant of the merger of the two holes.
This was already the subject of early studies using the Lazarus approach
\cite{Baker:2003ds, Campanelli:2004zw}. With the advent of the breakthroughs 
that allowed for longer accurate computations,
numerous empirical formulas relating the initial parameters 
$(q,\vec\chi_1,\vec\chi_2)$
(individual masses and spins) of the binary to those of the final
remnant $(m_f,\vec\chi_f,\vec{V}_f)$
have been proposed. These include formulas for the final
mass, spin, and recoil velocity
\cite{Barausse:2012qz,Rezzolla:2007rz,Hofmann:2016yih,Jimenez-Forteza:2016oae,Lousto:2009mf,Lousto:2013wta,Healy:2014yta,Zlochower:2015wga},
as well as algebraic properties of the final metric
\cite{Campanelli:2008dv,Owen:2010vw}. Recently,
the computation of the peak luminosity has also been the
subject of interest in relation to the observation of gravitational waves
\cite{TheLIGOScientific:2016wfe,TheLIGOScientific:2016pea,Healy:2016lce,Keitel:2016krm}.
The data in RIT catalog, along with the SXS~\cite{Mroue:2013xna} and
Georgia Tech.~\cite{Jani:2016wkt}, can be used by other groups to develop and
improve new empirical formulas for the remnant properties
and approximate/phenomenological waveform models
\cite{Babak:2016tgq,Khan:2015jqa}.

%%%%%%%%%%%%%%%%%%%%%%%%%%%%%%%%%%%%%%%%%%%%%
\acknowledgments The authors thank N.K.J-McDaniel, H. Nakano, and R.
O'Shaughnessy for discussions on this work.  The authors gratefully
acknowledge the NSF for financial support from NSF Grants No.
PHY-1607520, No. ACI-1550436, No. AST-1516150, No. ACI-1516125,  No.
PHY-1305730, No. PHY-1212426, No. PHY-1229173, No. AST-1028087, No.
PHY-0969855, No. OCI-0832606, and No. DRL-1136221.  Computational
resources were provided by XSEDE allocation TG-PHY060027N, and by
NewHorizons and BlueSky Clusters at Rochester Institute of Technology,
which were supported by NSF grants No.  PHY-0722703, No. DMS-0820923,
No. AST-1028087, and No. PHY-1229173.
%%%%%%%%%%%%%%%%%%%%%%%%%%%%%%%%%%%%%%%%

%%%%%%%%%%%%%%%%%%%%%%%%%%%%%%%%%%%%%%%%%%%%

\bibliographystyle{apsrev4-1}
\bibliography{../../Bibtex/references}

\end{document}